\documentclass{iau}

\usepackage{amsmath}
\usepackage{graphicx}
\usepackage{multirow}

\begin{document}

\lefttitle{Dorigo Jones et al.}
\righttitle{Emulating the Global 21 cm Cosmology Signal}

\jnlPage{1}{5}
\jnlDoiYr{2025}
\doival{}
\aopheadtitle{To appear in: Proceedings IAU Symposium}
\editors{I. Liodakis \& V. Efthymiou, eds.}

\title{Emulating Global 21 cm Cosmology Observations from the Lunar Far Side\\
to Achieve Quick and Reliable Physical Constraints}

\author{J. Dorigo Jones$^1$, J. O. Burns$^1$, D. Rapetti$^{2,3,1}$, Shah Mohammad Bahauddin$^{4,5}$, B. Reyes$^6$, and D. W. Barker$^1$}
\affiliation{$^1$Center for Astrophysics and Space Astronomy, Department of Astrophysical and Planetary Sciences, University of Colorado Boulder, CO 80309, USA \\ email: {\tt johnny.dorigojones@colorado.edu} \\
$^2$NASA Ames Research Center, Moffett Field, CA 94035, USA\\
$^3$Research Institute for Advanced Computer Science, Universities Space Research Association, Washington, DC 20024, USA\\
$^4$Laboratory for Atmospheric and Space Physics, University of Colorado, Boulder, CO 80303, USA\\
$^5$Center for Astronomy, Space Science and Astrophysics, Independent University, Bangladesh, Dhaka 1229, Bangladesh\\
$^6$University of Colorado Research Computing, University of Colorado Boulder, CO 80309, USA}

\begin{abstract} 
Efforts are underway to measure the global 21 cm signal from neutral hydrogen, which is a powerful probe of the early universe, using NASA radio telescopes on the far side of the Moon. Physics-based models of the signal are computationally expensive to perform Bayesian multi-parameter inferences, for which we have developed novel, publicly-available neural network emulators utilizing a Long Short-Term Memory (LSTM) network and a Kolmogorov-Arnold Network (KAN). {\sc 21cmLSTM} is currently the most accurate emulator in the community by leveraging the signal’s temporally-correlated structure, and {\sc 21cmKAN} maintains similar accuracy while training 75 times faster, by learning expressive functional transformations. Each emulator can fit realistic mock signals and obtain unbiased physical parameter constraints, with {\sc 21cmKAN} able to complete end-to-end training and inference in under 30 minutes. The implementation of machine learning tools like these in data analysis pipelines is important to fully exploit upcoming measurements of the cosmological 21 cm signal.
\end{abstract}


\maketitle

\section{Introduction}
The redshifted 21 cm cosmological signal emitted by neutral hydrogen tells a detailed story of the early universe, from the Dark Ages to the Cosmic Dawn and Epoch of Reionization. For decades, low radio frequency experiments have tried to measure the 21 cm brightness temperature relative to the radiation background to gain an understanding of the first stars and galaxies and the intergalactic medium that is inaccessible to telescopes operating at shorter wavelengths (see \citealt{Furlanetto06} for a review). Of particular interest is the 1D isotropic, or global, 21 cm signal, which is observed at $\nu \lesssim 235$ MHz, corresponding to redshifts $z\gtrsim5$. There is still no clear detection of the global 21 cm signal, partly because all experiments so far have been ground-based  radio telescopes (e.g., \citealt{EDGES, SARAS3, REACH}), which suffer from Earth's severe ionospheric distortion and radio frequency interference. Efforts are underway to measure the global 21 cm signal from the most radio quiet environment in the Solar System -- the far side of the Moon -- using for example LuSEE-Night \citep{Bale23} and \textit{FarView} \citep{FarView}.

To constrain the signal and the many astrophysical and cosmological parameters that shape its evolution, robust software and modeling tools are needed to extract the signal from the bright Milky Way foreground and ultimately complete full Bayesian inference analyses. Semi-numerical and semi-analytical simulations of the 21 cm signal (e.g., \citealt{Mesinger11, Fialkov14, Mirocha17}) are faster than fully hydrodynamic codes; however, it is highly costly to use them in inference pipelines that require evaluating the model across multidimensional parameter spaces to sufficiently fit a measurement of the signal. As a result, artificial neural network (NN) emulators of the 21 cm signal have been created to mimic such cosmological simulations in milliseconds via nonlinear regression and thus efficiently sample parameter posterior distributions (e.g., \citealt{REACH, Bevins24}), assuming systematic effects are properly addressed (e.g., \citealt{PaperII, Saxena23a}).

\section{Designing Novel Emulators to Enhance 21 cm Cosmology Inference Pipelines}
There are multiple surrogate models, or emulators, of the global 21 cm signal brightness temperature that offer varying amounts of emulation accuracy and speed (\citealt{Cohen20, globalemu, 21cmVAE, DorigoJones24}). We recently developed and released two novel emulators that harness NN architectures that leverage characteristics of the global 21 cm signal to achieve exceptional accuracy, training speed, and evaluation speed compared to other emulators. {\sc 21cmLSTM}\footnote[1]{\label{LSTM}\url{https://github.com/jdorigojones/21cmLSTM}} \citep{DorigoJones24} provides the most accurate emulations of the global 21 cm signal to date as a result of its inherent capability to capture the temporal evolution of the 21 cm brightness temperature across redshifts or low radio frequencies. This comes at a cost of training time, though, because of the sequential nature of recurrent NNs and the number of trainable parameters (i.e., weights and biases) required. As a result, we created {\sc 21cmKAN}\footnote[2]{\label{21cmKAN}\url{https://github.com/jdorigojones/21cmKAN}} \citep{DorigoJones25}, which has similar emulation accuracy as {\sc 21cmLSTM} but trains 75 times faster and evaluates 5 times faster than {\sc 21cmLSTM}, enabling efficient Bayesian parameter estimation analyses of different models and parameterizations.

We explored NN architectures that are designed to take advantage of properties inherent to the global 21 cm signal in order to realize state-of-the-art emulation. {\sc 21cmLSTM} is made of a Long Short-Term Memory (LSTM) network, which is a type of recurrent architecture that is specifically made for modeling long-term patterns in sequential data \citep{HochreiterSchmidhuber97, Gers00}. The global 21 cm signal evolves from high to low redshift, meaning that information is correlated across adjacent frequency bins. Thus, we developed {\sc 21cmLSTM} to leverage this temporal correlation for signal emulation. LSTM NNs store and propagate information through time using gated mechanisms that allow them to retain memories and capture temporal dependencies in data. However, because of their sequential as opposed to feedforward nature and the significant number of weights required for the memory cells, LSTM NNs can be slower to train and often require more complex optimization.

\begin{figure}
    \centering
    \includegraphics[width=1\linewidth]{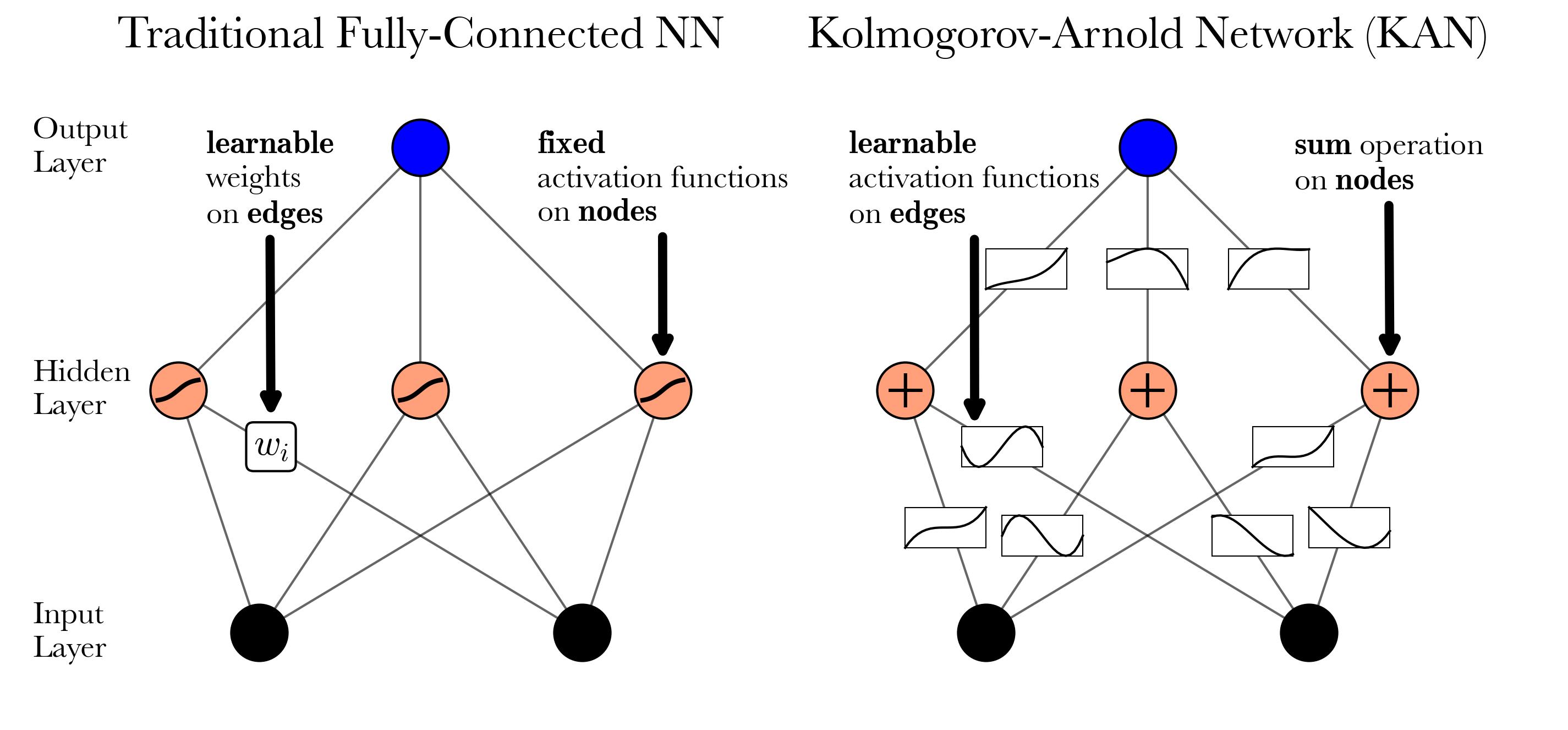}
    \caption{As shown in \citet{DorigoJones25}, these are architecture diagrams of a single-hidden-layer traditional fully-connected NN (left) and Kolmogorov-Arnold Network (right). In the KAN, activation functions are learned and applied to parameters on the edge connections between nodes and summed at the nodes. For traditional fully-connected NNs, the activations are pre-determined and fixed on the nodes, the scalar weights of which are learned on the edges.}
    \label{fig:KAN}
\end{figure}

\begin{figure}
    \begin{minipage}[c]{0.55\textwidth}
    \includegraphics[width=\linewidth]{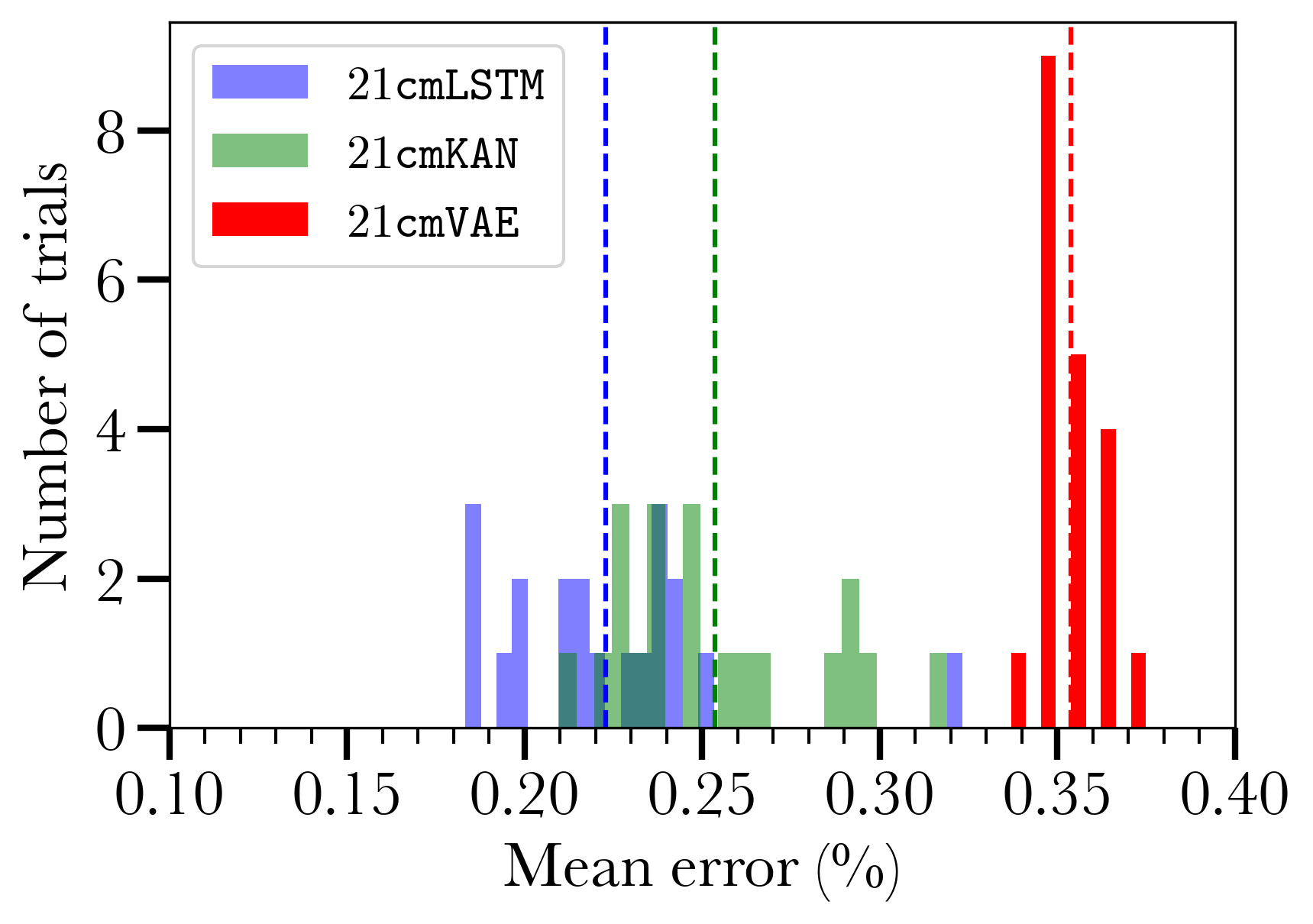}
    \end{minipage}\hfill
    \begin{minipage}[c]{0.44\textwidth}
    \caption{From \citet{DorigoJones25}, this is a histogram of the mean relative root mean square error for 20 trials of {\sc 21cmKAN} (in green) trained and tested on the {\sc 21cmGEM} data set.~The blue histogram is the error for 20 trials of {\sc 21cmLSTM} trained and tested on the same data \citep{DorigoJones24}. The red histogram is the approximate error for 20 trials of {\sc 21cmVAE} trained and tested on the same data (adapted from Figure 6 of \citealt{21cmVAE}).~Dashed lines depict the average emulation errors.}
    \label{fig:21cmkan}
    \end{minipage}
\end{figure}

We found that the Kolmogorov-Arnold Network (KAN\footnote[3]{\label{KAN}\url{https://github.com/KindXiaoming/pykan}}; \citealt{KAN}) can model the global 21 cm signal to a similar degree of accuracy as the LSTM NN but much more efficiently because of its simple and expressive architecture. KANs learn data-driven functional transformations (i.e., activation functions) to model complex relationships in data, rather than using static functions as do traditional fully-connected NNs. In contrast to LSTM NNs, KANs model the shape of the signal by directly learning the underlying 1D functional transformations on network edges and their summations at the nodes (see Figure~\ref{fig:KAN}), without the need for memory gates or recurrent structures. KANs are therefore well suited for multivariate curve emulation where preserving the sequential structure is important. With the help of optimized methods\footnote[4]{\label{EfKAN}\url{https://github.com/Blealtan/efficient-kan}}, training KANs is often faster and convergence is robust, particularly for low-dimensional problems involving functional compositions. This includes the global 21 cm signal and many other summary statistics found in physical sciences. Furthermore, since each transformation is explicitly represented as a learned function, the internal representations of KANs are easier to analyze, interpret, and verify compared to the scalar weights of fixed activation functions learned by traditional fully-connected NNs. In \citet{DorigoJones25}, we demonstrated that, while {\sc 21cmLSTM} achieved the highest overall accuracy among conventional neural architectures, {\sc 21cmKAN} attains comparable performance, as shown in Figure~\ref{fig:21cmkan}, with significantly faster convergence and reduced architectural complexity.

\begin{figure}
    \includegraphics[scale=0.265]{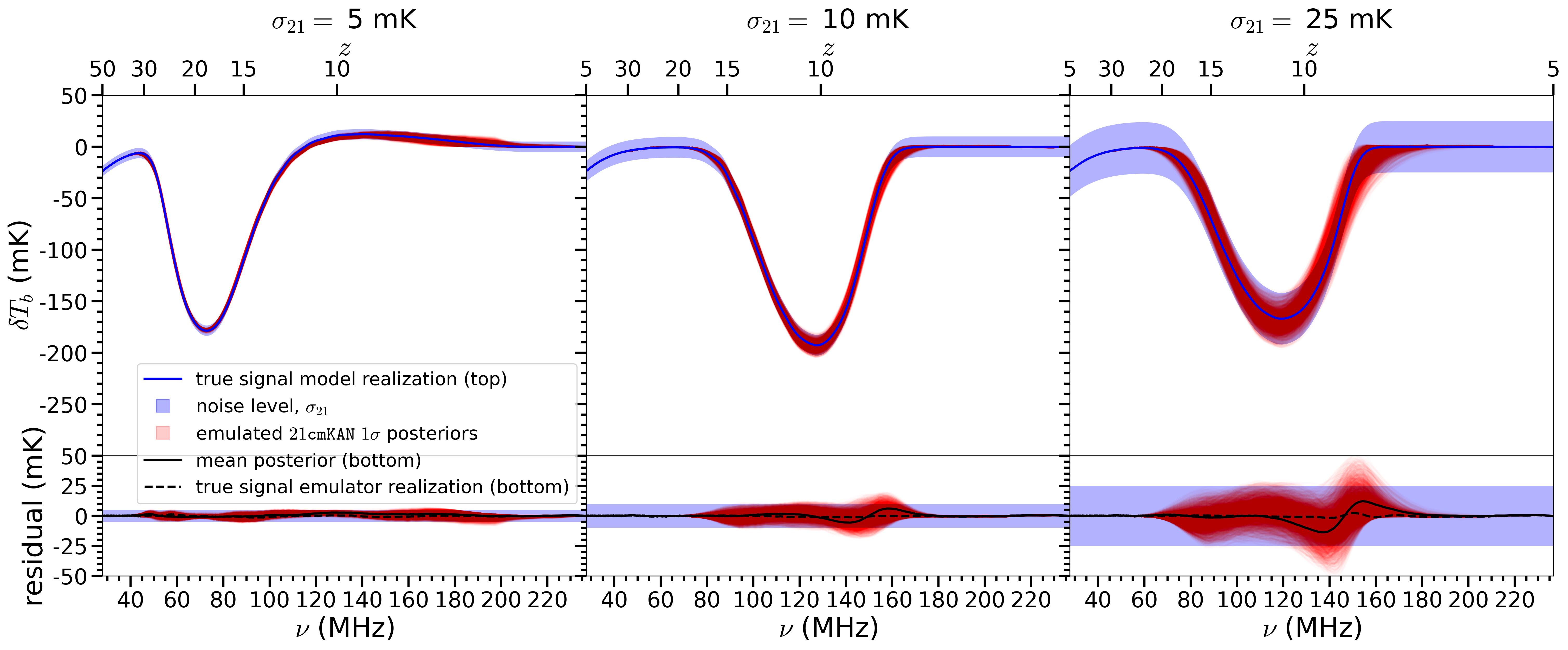}
    \caption{From \citet{DorigoJones25}. {\it Top:} Signal realizations of the 68\% best (i.e., $1\sigma$) posteriors (red) obtained from nested sampling analyses using {\sc 21cmKAN} to fit three global 21 cm signals (dark blue) from the {\sc 21cmGEM} test set \citep{21cmVAE} with added noise (light blue bands) of 5 mK (left), 10 mK (middle), and 25 mK (right). {\it Bottom:} Residuals between the corresponding true signal and each {\sc 21cmKAN} $1\sigma$ posterior (red), 
    the mean posterior (solid black), and the signal emulation (dashed black).}
    \label{fig:posterior}
\end{figure}

The combination of enhanced speed and accuracy facilitated by {\sc 21cmKAN} enables rapid and highly accurate physical parameter estimation analyses of multiple 21 cm models, which is needed to fully characterize the complex feature space across cosmological simulations and produce robust constraints on the early universe. {\sc 21cmKAN} can predict a given signal for two well-known models in the community in 3.7 ms on average and train in only 10 minutes, when utilizing a typical A100 GPU, achieving these speeds because of its expressive transformations and its relatively small number of trainable parameters compared to a memory-based emulator. In \citet{DorigoJones25}, we showed that {\sc 21cmKAN} required less than 30 minutes to train and fit simulated signals with added observational noise and obtain unbiased physical parameter posterior distributions (see posterior signal realizations in Figure~\ref{fig:posterior}). In addition, we found that the transparent architecture of {\sc 21cmKAN} allows the user to conveniently interpret and further validate its emulation results in terms of the sensitivity of the 21 cm signal to each physical model parameter. {\sc 21cmKAN} demonstrates the effectiveness of KANs and their ability to more quickly and accurately mimic expensive physical simulations or PDE solvers of the 21 cm signal in comparison to other types of NNs.

\section{Conclusions}
{\sc 21cmLSTM} and {\sc 21cmKAN} can be trained to emulate any data set of global 21 cm signals and subsequently used in Bayesian analyses to fit an observed signal and efficiently obtain unbiased physical parameter constraints to learn key properties of the early universe. The speed--accuracy combination of {\sc 21cmKAN} is particularly beneficial to the community because different emulator models will need to be trained to constrain the cosmological and astrophysical parameters across the different existing models of the global 21 cm signal. This will thus facilitate exploring posterior distributions of 21 cm signal parameters (e.g., \citealt{Qin20, REACH, DorigoJones23}) for different models to robustly fit and exploit measurements of the signal. For upcoming observations, {\sc 21cmLSTM} offers unprecedented accuracy when desired, and {\sc 21cmKAN} eliminates emulator training as a bottleneck in comprehensive inference pipelines. The adaptable and publicly-available nature of {\sc 21cmLSTM} and {\sc 21cmKAN} makes them immediately useful and integrable for the community.

We acknowledge support by NASA APRA grant award 80NSSC23K0013, a subcontract from UC Berkeley (NASA award 80MSFC23CA015) to the University of Colorado (subcontract \#00011385) for science investigations involving the LuSEE-Night lunar far side mission, and NASA award 80NSSC22K1264 to support radio astrophysics from the Moon.

\bibliography{dj25}{}
\bibliographystyle{aasjournalv7}
\end{document}